\documentclass[conference]{IEEEtran}
\IEEEoverridecommandlockouts
\usepackage{cite}
\usepackage{amsmath,amssymb,amsfonts}
\usepackage{algorithmic}
\usepackage{graphicx}
\usepackage{textcomp}
\usepackage{color}
\usepackage{epstopdf}
\usepackage{listings}
\usepackage{float}
\usepackage{color} %red, green, blue, yellow, cyan, magenta, black, white
\definecolor{mygreen}{RGB}{28,172,0} % color values Red, Green, Blue
\definecolor{mylilas}{RGB}{170,55,241}
\usepackage{float} % To fix the figure
\usepackage[center]{caption} %To put caption in center in figure or table
\usepackage{subcaption}
\usepackage{multicol}% http://ctan.org/pkg/multicols %used to settle long figure in between from top of the pages  
\usepackage{mathtools} %For prescript
\usepackage{cuted} % can temporarily leave two columns mode with its strip environment (begin{strip})
\usepackage[ruled,vlined]{algorithm2e}
\usepackage{hyperref}
\title{Dynamics of a Towed Cable with Sensor-Array for Underwater Target Motion Analysis}
\begin{document}

\author{
  \IEEEauthorblockN{ Rohit Kumar Singh}
  \IEEEauthorblockA{\textit{Department of Electrical Engineering} \\
    \textit{Indian Institute of Technology Patna}\\
    Bihar, India \\
    rohit\_1921ee19@iitp.ac.in}
  \and
  \IEEEauthorblockN{ Subrata Kumar}
  \IEEEauthorblockA{\textit{Department of Mechanical Engineering} \\
    \textit{Indian Institute of Technology Patna}\\
    Bihar, India \\
    subrata@iitp.ac.in}
  \and
  \IEEEauthorblockN{ Shovan Bhaumik}
  \IEEEauthorblockA{\textit{Department of Electrical Engineering} \\
    \textit{Indian Institute of Technology Patna}\\
    Bihar, India \\
    shovan.bhaumik@iitp.ac.in}
 
}
%\author{
%\IEEEauthorblockN{Rohit Kumar Singh\IEEEauthorrefmark{1}, Subrata Kumar \IEEEauthorrefmark{2}, Shovan Bhaumik \IEEEauthorrefmark{3}}\\
%\IEEEauthorblockA{\IEEEauthorrefmark{1}\IEEEauthorrefmark{2}\IEEEauthorrefmark{3}Department of Electrical Engineering, Indian Institute of Technology Patna, India}
%\IEEEauthorblockA{\IEEEauthorrefmark{1} \emph{rohit\_1921ee19@iitp.ac.in,}\IEEEauthorrefmark{2} \emph{subrata@iitp.ac.in,}\IEEEauthorrefmark{3} \emph{shovan.bhaumik@iitp.ac.in}}
%\vspace{0.05in}
%}
\maketitle

\begin{abstract}
During a war situation, many times an underwater target motion analysis (TMA) is performed using bearing-only measurements, obtained from a sensor array, which is towed by an own-ship with the help of a connected cable. It is well known that the own-ship is required to perform a manoeuvre in order to make the system observable and localise the target successfully. During the maneuver, it is important to know the location of the sensor array with respect to the own-ship. This paper develops a dynamic model of a cable-sensor array system to localise the sensor array, which is towed behind a sea-surface vessel. We adopt a lumped-mass approach to represent the towed cable. The discretized cable elements are modelled as an interconnected rigid body, kinematically related to one another. The governing equations are derived by balancing the moments acting on each node. The derived dynamics are solved simultaneously for all the nodes to determine the orientation of the cable and sensor array. The position of the sensor array obtained from this proposed model will further be used by TMA algorithms to enhance the accuracy of the tracking system.

\end{abstract}
\section{Introduction}

%Due to the sounds produced by the surface vessels as they move through water or noise from internal systems in operation, passive SONAR faces challenges in maintaining clarity among the internal and external sounds. To address this and facilitate uninterrupted listening to distant sources, passive SONAR is utilized as a towed cable sensor-array system (TCSAS) \cite{abraham2019underwater}. This system involves a series of hydrophones towed behind a submarine or surface ship on a long, slender, flexible cylindrical cable extending for kilometers. This configuration ensures that the array's sensors remain distant from the ship's inherent noise sources \cite{yang2023dynamic} as shown in Fig. \ref{fig1}.

\begin{figure}[htbp]
				\includegraphics[scale = 0.47]{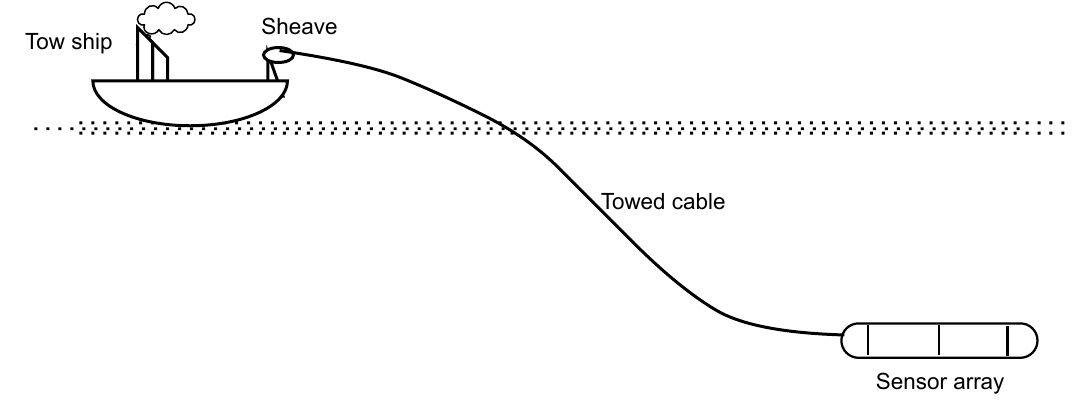}
				\caption{ Towed cable sensor-array system.}
				\label{fig1}
			\end{figure}

Target motion analysis (TMA) is used to track and predict the motion of a target, such as enemy submarines or surface ships \cite{radhakrishnan2018gaussian}. Often, a passive TMA is preferred due to some tactical advantages and it uses passive SONAR to collect acoustic signals emitted by the target. A passive SONAR operates by listening without transmitting, and it can either be mounted on a hull or towed behind a ship \cite{lemon2004towed}. If the sensor array is not hull mounted, it is attached to a long cable that is towed by the ship \cite{abraham2019underwater} in order to minimize the interference of ship's sound during sensing \cite{abraham2019underwater}. This configuration ensures that the sensor array remains distant from the ship's inherent noise sources \cite{yang2023dynamic} as shown in Fig. \ref{fig1}.

The Kalman filtering techniques are then applied to predict and update the target's state based on the collected measurements \cite{payan2021passive}. In the passive mode, SONAR can only measure bearing angle \cite{stiles2013dynamic}. Tracking a target with only bearing measurement is known as bearing-only tracking (BOT). The BOT is indeed a challenging task, and the own-ship moving with constant velocity needs to execute a manoeuvre to ensure the observability of the tracking system \cite{northardt2022observability} so that the tracking filter tracks the target and the estimation converges \cite{singh2022passive}. 

To track a target, filters require an accurate location of the sensor array \emph{w.r.t.} ship \cite{yang2013dynamic}. It is easily obtained if the own-ship maintains a steady, level course during measurement sampling intervals. However, any manoeuvres or changes in the course of the ship can disrupt the path of the sensor-array, and its position becomes untraceable \cite{zhang2023dynamic}. So, a dynamic model of towed array system is warranted which will provide sensor location accurately for all sorts of maneuvers of the ship. In this paper, we tried to develop a dynamic model using Newtonian mechanics which when solved will provide the position of the sensor array at any point in time.

Research on the configuration of towed cable-array systems (in different contexts) traces its origins back to the 1950s and has witnessed substantial development over the last few decades \cite{lemon2004towed}. This evolution can be broadly categorized into two aspects: a simulation-based approach \cite{yuan2016research}, and experimental observation \cite{jung2002numerical,liu2013transient,chen2016experimental}.
In the simulation-based approach, the towed cable is conceptualized as a flexible body, while the sensor array is treated as rigid \cite{zhu2003dynamic}. The flexible modelling of the cable involves discretizing it into numerous small finite elements, where the mass of each element is concentrated into a node \cite{sun2011dynamic}. This approach is also known as the lumped mass method.  
%There are four commonly existing methods for modelling the solution of the dynamics of a towed system: (i) lumped mass method \cite{calnan2018reference,driscoll2000development,du2019numerical,yang2013dynamic}, (ii) finite element method \cite{buckham2004development,rodriguez2020simulation,quan2015geometrically,sun2011dynamic}, (iii) finite difference method \cite{wang2015parameters,yuan2014finite}, and (iv) slender/rod flexible segment method \cite{kamman1985modelling,xu2015robust}. Each method offers unique advantages and limitations depending on the specific application and requirements.
The lumped mass method \cite{calnan2018reference,driscoll2000development,du2019numerical,yang2013dynamic} uses Newton's second law of motion to model the dynamic behaviour of towed cable. It solves the resulting non-linear ordinary differential equation using finite difference approximations, transforming the continuous problem into a discrete one \cite{huang1994dynamic}.
%The Finite Element Method (FEM) divides the continuous towed cable into discrete elements, allowing for different geometrical and material properties within each element. Although individual elements may vary, but they share the same governing mathematical equations. By assembling these elements, the FEM enables the algorithmic modelling of complex cable geometries, allowing for varying properties along the cable length \cite{zhu2003dynamic}. This method calculates the relative displacement of the cable, determining the new position by adding the relative displacement to the previous one \cite{sun2011dynamic}.
%The finite-difference method transforms the governing difference equations of cable dynamics into a system of algebraic linear approximations, which are solved using matrix algebra methods \cite{ablow1983numerical}.
%The Flexible Segment Method models the cable by dividing it into flexible segments, each representing a flexible element interconnected at nodes. These segments undergo significant longitudinal and lateral deformations. The dynamic behaviour is described by a set of nonlinear partial differential equations derived from the deformation theories applied to these flexible elements. Numerical integration is then utilized to simulate the dynamic response of the towed cable \cite{huston1982validation}.
All these existing works modelled the marine towing cable for different applications such as marine seismic exploration for oil and gas detection \cite{guo2021numerical}, towed submarine for mineral exploration \cite{yang2023dynamic}, sea-terrain mapping \cite{zhao2021numerical}, establishing the communication relay with the onshore station \cite{feng2022study}, and offshore energy exploration \cite{liu2023study}. To best of our knowledge, none of the existing work has performed the dynamics modelling of the towed cable sensor-array system (TCSAS) which could be useful for the purpose of underwater TMA.

In this study, we have adopted the first physics principles, such as Newton's laws of motion and principles of conservation of momentum, to derive the governing equations for the dynamics modelling of the towed cable sensor-array system (TCSAS) without relying on empirical data. This approach offers a more rigorous and comprehensive understanding of the system's underlying physics.
This study uses a lumped-mass model to represent the towed cable as a flexible body with discretized cable elements serving as interconnected rigid bodies, kinematically related to one another \cite{meriam2020engineering}. The connecting flexible cable is assumed to consist of two interconnected rigid bodies, and the sensor array is hinged at the end of the second segment. The governing equations account for gravity, buoyancy, lift, towed force, reaction force, material properties, and hydrodynamics acting on the lumped mass point and they are derived by balancing moments at each node, which are solved simultaneously for all the nodes at every instant to determine the orientation of each discretized cable element and sensor array.  The proposed model is implemented to simulate the dynamics of the towed sensor-array for an engagement scenario involving a target and an own-ship manoeuvre \cite{radhakrishnan2018gaussian} and the simulation results are presented in the paper. 

%The coordinates of the center of gravity (CG) of the sensor array obtained from this model are fed into the state estimation algorithm. 

%\textcolor{blue}{The dynamics modelling in this work is distinguished from the existing work by deriving the model directly from the first principles technique rooted in fundamental physics, potentially offering a more rigorous and comprehensive understanding of the system's underlying physics \cite{meriam2020engineering}}.
%This work distinguishes itself from existing research work in which the numerical algorithm technique utilized to solve the cable dynamics is overemphasized. 
% This investigation develops a numerical approach based on the lumped mass model to effectively simulate the dynamics of a towed cable sensor array, with the consideration of effects arising due to reaction force at the interconnected nodes. 

% The paper is arranged as follows: Section II develops the mathematical model for the dynamics of TCSAS, presenting a resulting non-linear ordinary differential equation that is solved using numerical methods. Section III then implements the proposed model to simulate the motion of the towed cable array, with concluding remarks provided in Section IV. 

	\section{Mathematical Model}

 Fig. \ref{fig1} shows a surface vessel towing a sensor-array which is connected through a flexible cable.  This section aims to derive a mathematical model to provide the location of towed sensor array when the ship's motion is known. The dynamics modelling of the towed cable sensor-array system is accomplished using the fundamental laws of Newtonian mechanics, by considering the following assumptions: (i) the sensor array is hinged at the end of cable, (ii) the cable is a flexible body, (iii) ship motion is independent of the towed cable senor-array because the ship's mass is several order of magnitude higher than the cable sensor-array system, (iv) water surface is still, (v) the deformation of cable is negligible under the influence of flowing water, (vi) and drag coefficient is constant irrespective of angle of attack between the cable-array system and fluid surface.
 
    %Fig. \ref{fig1} shows a surface vessel towing a cylindrical array of sensor which is connected through a flexible cable.
   % The flexible cable is to be modelled.
    %The rigid body can be modelled as flexible body by discretising the body and their dynamics is obtained by applying the moment balance equation at the junction point whose motions are related kinematically.
 The flexible rope is modelled by discretizing a single rigid body into multiple interconnected rigid elements, with each element's mass concentrated into a node. Fig. \ref{fig2} illustrates the free body diagram, depicting a rope connected to the ship through a spherical joint at point A, and a rigid sensor array segment is hinged to the end of the rope at point B. The tow point A experiences a towing force from the surface ship.  Additional forces, such as gravity, buoyancy, lift, reaction force, and hydrodynamics, act on the towed body. The dynamics of TCSAS are obtained using the moment balance equation at each node, and they depend on various factors, including the forces acting on the towed body, the number of segments used for modelling the rigid body, types of joints, the medium of motion, towed body structure, and its material composition.
    %These force is transferred to the rest of the body, affecting its dynamics. Apart from the applied motion from the surface ship, the dynamics of the towed body depend on factors such as the number of segments/joints used for modelling the rigid body, types of joints, body structure, medium of motion and material composition of body. 
    %The dynamics of towed system is obtained using the moment balance equation at each junction.
   % The sum of moments about the junction point due to all the external forces must be equal to the moments of resultants. 
     \subsection{General/Governing equation of motions}
    \begin{figure}
				\includegraphics[scale = 0.4]{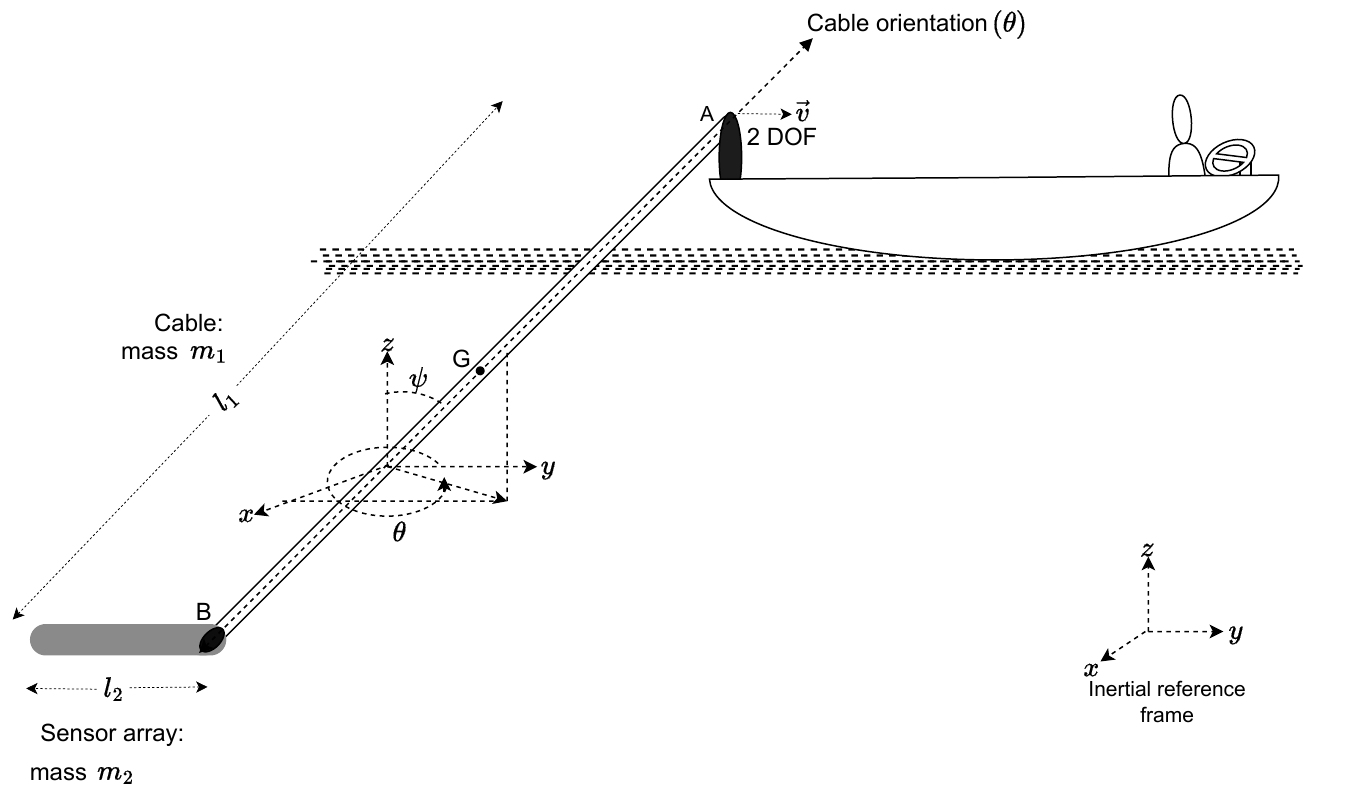}
				\caption{Free body diagram representation of a rigid body hinged to a ship.}
				\label{fig2}
			\end{figure}
  %  The general equation of moments about the point A is given as sum of moments of forces external to the system, since the internal forces cancels each other, and their summation is zero.
  %  Thus the moment sum is
  %  \begin{equation}
  %  	\textstyle\sum{M_{A}} =\dot{H}_{G} + \vec{\rho} \times m\vec{a_{A}}
   % \end{equation}
    %The moment equation can be easily visualised with the aid of Fig. \ref{fig2}, where the $\dot{H}_{G}$ is the sum of all the external moments about the point G, and $\vec{\rho} \times m\vec{a_{A}}$ is transfer of moment from center of mass to point A.  
    %%%%%%%%%%
    The angular momentum about point A, visualised with the aid of Fig. \ref{fig2} is 
    \begin{equation}
    	L_A =L_{G} + \vec{r}_{AG} \times m\vec{v}, 
    \end{equation}
 where $L_G$ is the angular momentum about the centre of gravity (CG) $G$, $\vec{r}_{AG}$ is a vector from towing point $A$ to CG of the cable, $G$, and $m\vec{v}$ is the linear momentum acting at G.
    %The above equation states that the absolute angular momentum about point A equals the angular momentum about G plus the moment about A of the linear momentum $m\vec{v}$ of the system considered concentrated at $G$. 
The resultant of all the external forces acting on the system is represented as resultant force $\textstyle\textstyle\sum F$ through $G$ and its corresponding moment is $\textstyle\sum M_G$. The sum of moments about point $A$ of Fig. \ref{fig2}, due to all the forces external to the system, must be equal to the moment of their resultants, which can be expressed as
      \begin{equation}
    	\textstyle\textstyle\textstyle\sum{M_A} =\dot{H}_{G} + \vec{r}_{AG} \times m\vec{a}_A,
    \end{equation}
where $\dot{H}_{G} = I\Ddot{\theta}$ is the moment due to the resultant force at $G$, $I$ is the moment of inertia, and $\Ddot{\theta}$ is the angular acceleration. $\textstyle\textstyle\sum{M_{A}}$ is the sum of external moments about point $A$. The motion of the moving coordinate system in Newtonian mechanics is specified with respect to the inertial frame of reference, as shown in Fig. \ref{fig2}.
   
\begin{figure}
				\includegraphics[scale = 0.43]{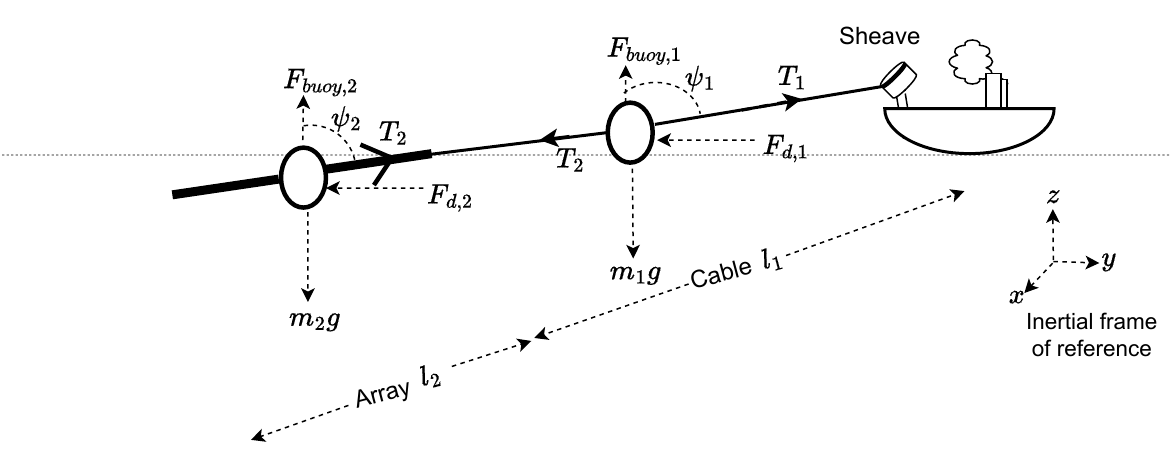}
				\caption{Lumped mass model for TCSAS in steady state condition.}
				\label{figss}
			\end{figure}

The TCSAS achieves a steady-state equilibrium configuration when the surface ship maintains a stable operating speed \cite{guo2021numerical}. The depth of the towed body at this configuration is determined using the tension balance equation at the tow point. The ship is moving with constant velocity so the tension balance is performed in an inertial reference frame as illustrated in Fig. \ref{figss}. The pitch angle orientation of the towed system under steady state condition is $\psi_1$. 
    The sensor-array is acted upon by the following forces: a vertical force $m_2g$ due to gravity, a horizontal force $F_{d,2}$ due to drag, a buoyant force $F_{buoy,2}$, and a tension $T_2$ exerted by cable. The tension force on sensor-array is at an angle of $\psi_2$ away from the vertical axis. The force balance equation is 
    \begin{equation} \label{eqdepth1}
        \begin{split}
            m_2g =& T_2\cos\psi_2-F_{buoy,2},\\
            F_{d,2} = & T_2 \sin\psi_2.
        \end{split}
    \end{equation}
   The following forces act on the cable: a vertical force of $m_1g$ due to gravity, a horizontal force $F_{d,1}$ due to drag, a buoyant force $F_{buoy,1}$, a tension force $T_1$ exerted by ship, and a tension force $-T_2$ exerted by sensor-array. The tension force on the cable is at an angle of $\psi_1$ away from the vertical axis. The force balance equation for the rope is 
    \begin{equation} \label{eqdepth2}
        \begin{split}
            m_1g +T_2cos\psi_2 =& T_1\cos\psi_1-F_{buoy,1}\\
            F_{d,1}+T_2\sin\psi_2 = & T_1 \sin\psi_1.
        \end{split}
    \end{equation}
 Using Eqn. \eqref{eqdepth1} and \eqref{eqdepth2}, we get a steady state pitch angle orientation of cable-array system \emph{w.r.t.} to the z-axis, given as
 \begin{equation} \label{eqdepth}
     \tan\psi_1 = \dfrac{F_{d,1}+F_{d,2}}{m_1g+m_2g+F_{buoy,1}+F_{buoy,2}}.
 \end{equation}
  The depth control of a conventional towed system can be achieved by adjusting either the towing speed or the physical parameters of the towing cable or the towed sensor array \cite{feng2022study}.
    %%%%%%
    \subsection{Moment balance equation for a single rigid body}
    \begin{figure}
				\includegraphics[scale = 0.5]{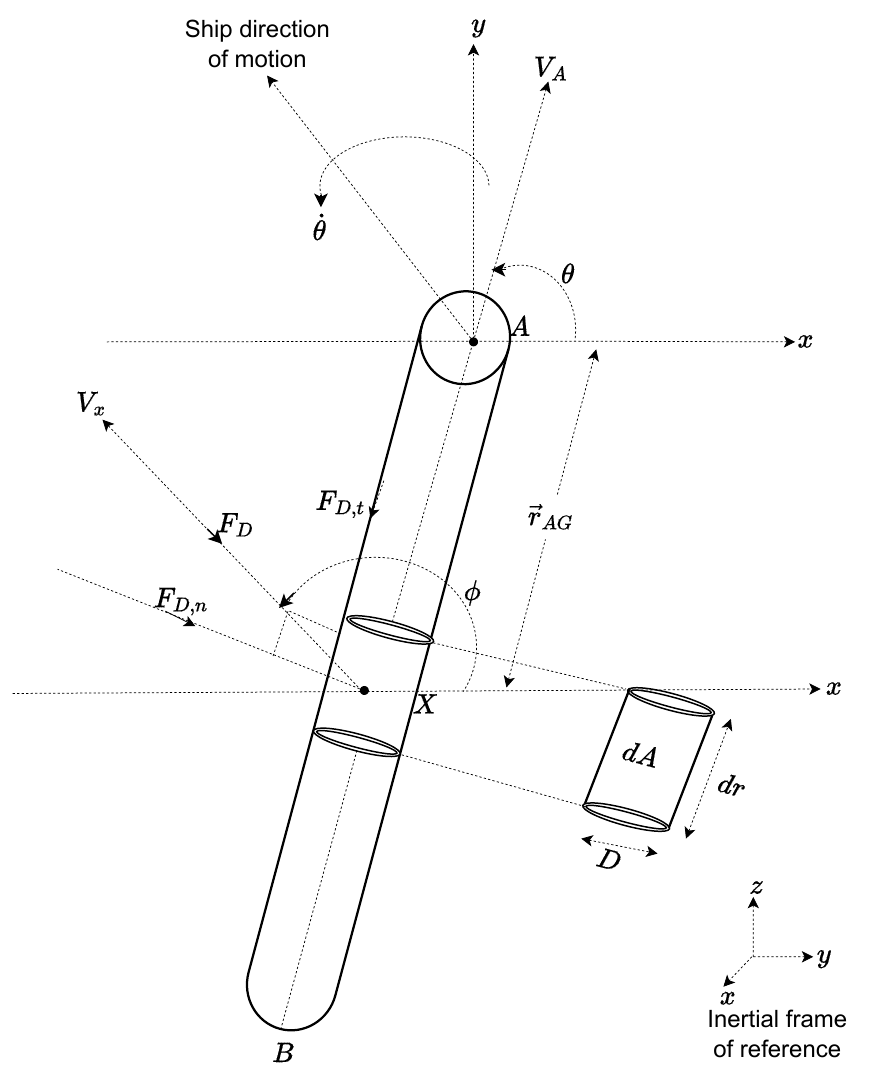}
				\caption{Free body diagram of a single body towed to ship.}
				\label{drag}
			\end{figure}
    Firstly, we derive the moment balance equation for a single rigid body while considering the hydrodynamics associated with it in $xy$ inertial frame of reference and then extend it for interconnected bodies.
    The illustration is shown in Fig. \ref{drag}, where a single rigid body AB is towed to the surface ship at point $A$. The moment balance equation at point $A$ is given as 
    \begin{equation}\label{momentbal1}
        \textstyle\textstyle\sum M_{e,A} =  \dot{H}_{G} + \vec{r}_{AG} \times m\vec{a}_A,
    \end{equation}
    where $\textstyle\sum M_{e,A}$ is the moments generated due to the sum of external forces, which in this case will be drag force. $\vec{r}_{AG}=\frac{L}{2} \cos\theta \hat{i} + \frac{L}{2} \sin\theta \hat{j} $ and $\vec{a}_A=\ddot{x}_{A} \hat{i} + \ddot{y}_{A} \hat{j}$. Substituting these in Eqn. \eqref{momentbal1}, the moment balance equation at point $A$ is 
     \begin{equation} \label{momentbal2}
        \textstyle\sum M_{d,A} = I\ddot{\theta}+\frac{ml}{2}[\ddot{y}_{A} \cos\theta - \ddot{x}_{A} \sin\theta ],
     \end{equation}
    where $\textstyle\sum M_{d,A}$ is the sum of moment due to drag force acting at point $A$. There are two components of drag force, \emph{i.e.} pressure drag ($F_{D,n}$) and shear friction drag ($F_{D,t}$) acting on the body AB along normal and tangential directions, respectively as shown in Fig. \ref{drag}. 
    
    It is to be noted that the $\textstyle\sum{M_{d,A}}$ is the moments generated due to the resultant of pressure drag force (normal drag) acting perpendicular to the rigid body motion. The drag force due to shear friction (tangential drag) won't generate any moment as it acts along the tangential direction to the motion of rigid body.
    
    Now we proceed to find the components of Eqn. \eqref{momentbal2} using free body diagram Fig. \ref{drag}. A fractional element $X$ of area $dA=Ddr$, with a velocity $\vec{V}_X$ at an angle $\phi$ \emph{w.r.t.} $x$-axis, is shown in Fig. \ref{drag}. The drag force $d\vec{F}_d$ (combined tangential and normal) acting on element $X$ is in the opposite direction to velocity $\vec{V}_X$. The normal component of $\vec{V}_X$ is $V_{X,n} = V_X sin(\phi-\theta)$.
     The moment generated at point $A$, due to normal drag force $d\vec{F}_{d,n}$ acting on fractional element $X$ is given as
    \begin{equation} \label{eqmoment1}
        M_{d,A} = \vec{r}\times d\Vec{F}_{d,n}, 
       \end{equation}
        where 
        \begin{equation} \begin{split}\label{eqdrag}
        d\vec{F}_{d,n} = & -\dfrac{1}{2} \rho C_{D,n} dA \vec{V}_{X,n} |\vec{V}_{X,n}|,\\
          = & -\dfrac{1}{2} \rho C_{D,n} D \vec{V}_{X} \sin(\phi-\theta) |\vec{V}_{X} \sin(\phi-\theta)| dr.
       \end{split}  \end{equation}
    In the above equation, $\vec{V}_X$ and $\phi$ are unknown and have to be substituted with the known quantities.
    %For that, we perform substitutions utilizing the free body diagram Fig. \ref{drag}. 
    The velocity $\vec{V}_X$ is the vector sum of its translational part $\vec{V}_A$ and the rotational part $\vec{V}_{X/A}$. Thus,
    \begin{equation} 
	\vec{V}_{X}  = \vec{V}_{A} + \vec{V}_{X/A}, \end{equation}
	where, \begin{equation} \vec{V}_{A} = \dot{x}_{A}\hat{i} + \dot{y}_{A}\hat{j} , \  \vec{V}_{X/A} = \vec{\theta} \times \vec{r}_{XA}, \ \vec{\dot{\theta}} = \dot{\theta}\hat{k}.\end{equation} 
 On substituting, we get
 \begin{equation} \begin{split} \label{eqvel1}
	\vec{V}_{X}	& = (\dot{x}_{A}\hat{i} + \dot{y}_{A}\hat{j}) + \dot{\theta}\hat{k} \times (r\cos\theta\hat{i}+r\sin\theta\hat{j}), \\
	&	= (\dot{x}_{A} -\dot{\theta}r \sin\theta )\hat{i} +(\dot{y}_{A} +\dot{\theta} r \cos \theta  )\hat{j},\end{split}
\end{equation} \begin{equation}\label{eqvel2}
	 \ |\vec{V}_{X}| = \sqrt{ \dot{x}_{A}^2 +\dot{y}_{A}^2+\dot{\theta}^2r^2+2\dot{\theta}r( \dot{x}_{A}\sin\theta-\dot{y}_{A}\cos\theta ) },
\end{equation}
\begin{equation}\label{eqvel3}
    \phi = \tan^{-1} (\dfrac{\dot{y}_{A} +\dot{\theta} r \cos \theta}{\dot{x}_{A} -\dot{\theta}r \sin\theta}).
\end{equation}
%\begin{equation} \label{eqsin} \begin{split}
  %  \sin(\theta-\phi)  =& \sin \theta \cos \phi - \cos \theta \sin \phi \\
  %             = & \dfrac{(\dot{x}_{A}-\dot{\theta}r\sin\theta)\sin\theta- (\dot{y}_{A}+\dot{\theta}r\cos\theta)\cos\theta}{\sqrt{\dot{x}_{A}^2+\dot{y}_{A}^2+\dot{\theta}^2 r^2 +2\dot{\theta}r(\dot{x}_{A}\sin\theta-\dot{y}_{A}\cos\theta)  }}.
  %  \end{split}\end{equation}
    Substituting Eqn. \eqref{eqvel1}, \eqref{eqvel2} and \eqref{eqvel3} into Eqn. \eqref{eqdrag} gives the normal drag force acting on fractional element $X$, given as
    \begin{equation} \label{eqdrag2} \begin{split}
        dF_{d,n}  
        %&  -\frac{1}{2}\rho C_{D,n} D  [(\dot{x}_{A}-\dot{\theta}r\sin\theta )\hat{i}+ (\dot{y}_{A}+\dot{\theta}r\cos\theta )\hat{j}] \\
       % & [\dot{x}_{A}\sin\theta-\dot{y}_A \cos\theta-\dot{\theta}r ]^2 [\dot{x}_{A}^2+\dot{y}_{A}^2+\dot{\theta}^2 r^2 +2\dot{\theta}r \\
        %& (\dot{x}_{A}\sin\theta-\dot{y}_{A}\cos\theta)]^{-\frac{1}{2}}   dr,\\
        = & -\frac{1}{2}\rho C_{D,n} D  [(\dot{x}_{A}-\dot{\theta}r\sin\theta )\hat{i}+ (\dot{y}_{A}+\dot{\theta}r\cos\theta )\hat{j}] \\
        & [\dot{x}_{A}\sin\theta-\dot{y}_A \cos\theta-\dot{\theta}r ]^2 (|\vec{V}_{X}(r)|)^{-1}    dr.
    \end{split}\end{equation}
  
    Similarly, the tangential drag force acting on fractional element $X$ is given as
    \begin{equation} \label{eqdrag3} \begin{split}
        dF_{d,t}  =&  -\frac{1}{2}\rho C_{D,t} \pi D  [(\dot{x}_{A}-\dot{\theta}r\sin\theta )\hat{i}+ (\dot{y}_{A}+\dot{\theta}r\cos\theta )\hat{j}] \\
        & [\dot{x}_{A}\cos\theta-\dot{y}_A\sin\theta ]^2 (|\vec{V}_{X}(r)|)^{-1}   dr.
    \end{split}    
    \end{equation}
 The moment generated at point $A$ due to normal drag force acting over the entire segment AB is given as
 \begin{equation}
     M_{d,A} =\int_{0}^{L} \vec{r} \times d\vec{F}_{d,n} .
\end{equation}
 Substituting $d\vec{F}_{d,n}$ from Eqn \eqref{eqdrag2}, we get
 \begin{equation} \label{eqmdrag}\begin{split}
     M_{d,A} = & -\frac{1}{2}\rho C_{D,n} D \int_{0}^{L} ( r\cos\theta\hat{i}+r\sin\theta\hat{j} ) \times [(\dot{x}_{A}- \\ & \dot{\theta}r\sin\theta )\hat{i}+ (\dot{y}_{A}+\dot{\theta}r\cos\theta )\hat{j}]  [\dot{x}_{A}\sin\theta- \dot{y}_A \cos\theta \\ 
     &-\dot{\theta}r ]^2(|\vec{V}_{X}(r)|)^{-1}   dr \\
     = & -\frac{1}{2}\rho C_{D,n} D \int_{0}^{L} r (\dot{y}_{A}\cos\theta-\dot{x}_{A}\sin\theta-\dot{\theta}r)^3 \\ & (|\vec{V}_{X}(r)|)^{-1}    dr.
 \end{split}     
 \end{equation}
   The moment balance equation at point $A$ using Eqns. \eqref{momentbal2} and \eqref{eqmdrag} is 
   \begin{equation} \begin{split} \label{thetadd}
     & \int_{0}^{L} M_{d,A} dr = I\ddot{\theta}_1+\frac{ml}{2}(\ddot{y}_{A} \cos \theta - \ddot{x}_{A} \sin \theta ),\\
     & -\frac{1}{2}\rho C_{D,n} D \int_{0}^{L} r (\dot{y}_{A}\cos\theta-\dot{x}_{A}\sin\theta-\dot{\theta}r)^3 (|\vec{V}_{X}(r)|)^{-1} dr\\& =I\ddot{\theta}_1+\frac{ml}{2}(\ddot{y}_{A} \cos \theta - \ddot{x}_{A} \sin \theta ).
   \end{split}\end{equation}

 The yaw orientation $\theta$ of body AB is obtained by solving the  $2^{nd}$ order ODE of $\ddot\theta$ from Eqn. \eqref{thetadd}. The kinematics equations given below use the yaw orientation to obtain the position, velocity and acceleration at the end of the segment AB \emph{i.e.} at point B.
 \begin{equation} \label{eqkinematics} \begin{split}
 	        x_{B} =& x_{A} + l \cos\theta_{1}, \ y_{B} = y_{A} + l \sin\theta_{1},\\
				\dot{x}_{B} =& \dot{x}_{A}-l\dot{\theta}_{1}\sin\theta, \ \dot{y}_{B} =\dot{y}_{A}+l\dot{\theta}_{1}\cos\theta, \\
				\ddot{x}_{B}=& \ddot{x}_{A}-l\ddot{\theta_{1}}\sin\theta_{1}-l\dot{\theta_{1}}^2\cos\theta_{1},\\
				\ddot{y}_{B}= & \ddot{y}_{A}+l\ddot{\theta_{1}}\cos\theta_{1}-l\dot{\theta_{1}}^2\sin\theta_{1}.\\
			\end{split}\end{equation}
   Similarly, the moment balance equation can be derived for each interconnected rigid segment, yielding a second-order ODE for yaw orientation at each node.
   \subsection{Modelling of cable-array system}
   \begin{figure}
				\includegraphics[scale = 0.5]{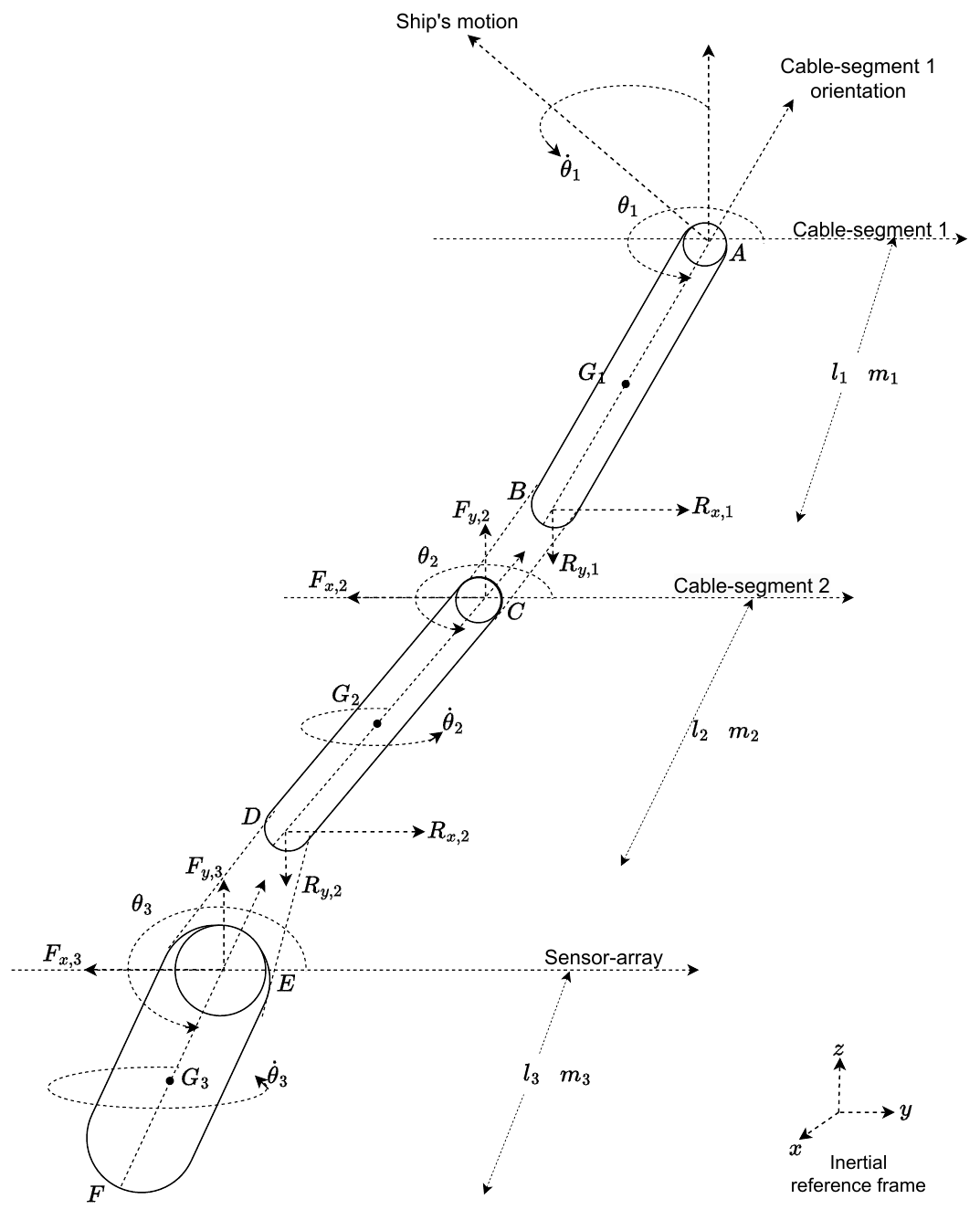}
				\caption{Free body diagram of a towed cable sensor-array system.}
				\label{cablearray}
			\end{figure}
   The cable is modelled by discretizing the rigid cable in two segments, and the array is hinged to the end of cable $2^{nd}$ segment as shown in free body diagram Fig. \ref{cablearray}.\\ 
  The moment balance equation at node $A$ for cable-segment 1 at time instant $k$ is
 \begin{equation}
     \textstyle\sum M_{e,A}^k =  I_1\ddot{\theta}_{1}^k+\frac{m_1l_1}{2}(\ddot{y}_{A}^k \cos\theta_{1}^k - \ddot{x}_{A}^k \sin\theta_{1}^k ),
 \end{equation}
 where $\theta_1^k$ is the orientation of the cable-segment 1 at $k^{th}$ instant, $(\ddot{x}_{A}^k,\ddot{y}_{A}^k)$ is acceleration along $(x,y)$ at node A which is obtained from ship input, $\textstyle\sum M_{e,A}^k$ is the sum of moment due to the drag force acting on segment AB and the moment due to the reaction force from the segment CD acting on segment AB at $k^{th}$ instant, given as
 \begin{equation} \begin{split}
     \textstyle\sum M_{d,A}^k-\textstyle\sum M_{rtn,A}^k =&  I_1\ddot{\theta}_{1}^k+\frac{m_1l_1}{2}(\ddot{y}_{A}^k \cos\theta_{1}^k - \ddot{x}_{A}^k  \sin\theta_{1}^k ),
 \end{split}\end{equation}
 where the moment generated due to reaction force at point $A$ is 
\begin{equation} 
  \textstyle\sum M_{rtn,A}^k =  R_{x,1}^k(l_1\sin{\theta}_{1}^k) - R_{y,1}^k(l_1\cos{\theta}_{1}^k).  
 \end{equation}
$(R_{x,1}^k,R_{y,1}^k)$ is the reaction force acting at node $B$ of cable-segment 1 which is generated due to cable-segment 2 and sensor-array along $(x,y)$ direction, respectively, expressed as
 \begin{equation} \label{eqreaction}\begin{split}
     R_{x,1}^k = & -(m_2\ddot{x}_{G2}^k -D_{n,x,2}^k - D_{t,x,2}^k + F_{x,3}^k), \\
     R_{y,1}^k = & -(m_2\ddot{y}_{G2}^k -D_{n,y,2}^k - D_{t,y,2}^k + F_{y,3}^k),
 \end{split} \end{equation}
 where $(m_2\ddot{x}_{G2}^k,m_2\ddot{y}_{G2}^k)$ are force components of cable-segment 2 along $(x,y)$ at $k^{th}$ instant, $D_{n,2}^k = (D_{n,x,2}^k, D_{n,y,2}^k)$ and $D_{t,2}^k = (D_{t,x,2}^k, D_{t,y,2}^k)$ are normal and tangential drag force along $(x,y)$, respectively acting on cable-segment 2 at $k^{th}$ instant, given as 
 \begin{equation} \label{eqdragcomp} \begin{split}
     D_{n,x,2}^k =&  -\frac{1}{2}\rho C_{D,n} D \int_{0}^{L} (\dot{x}_{C}^k-\dot{\theta}_2^kr\sin\theta_2^k )  (\dot{x}_{C}^k\sin\theta_2^k-\\
     &\dot{y}_C^k  \cos\theta_2^k -\dot{\theta}_2^k r )^2 (|\vec{V}_2^k(r)|)^{-1} dr,\\
    D_{n,y,2}^k = & -\frac{1}{2}\rho C_{D,n} D  \int_{0}^{L} (\dot{y}_{C}^k+\dot{\theta}_2^kr\cos\theta_2^k )(\dot{x}_{C}^k\sin\theta_2^k - \\
         & \dot{y}_C^k \cos\theta_2^k-\dot{\theta}_2^kr )^2 (|\vec{V}_2^k(r)|)^{-1}   dr,\\
    D_{t,x,2}^k = & -\frac{1}{2}\rho C_{D,t}\pi  D \int_{0}^{L} (\dot{x}_{C}^k-\dot{\theta}_2^kr\sin\theta_2^k )  (\dot{x}_{C}^k\cos\theta_2^k- \\
     &\dot{y}_C^k \sin\theta_2^k )^2   (|\vec{V}_2^k(r)|)^{-1}   dr,\\
     D_{t,y,2}^k =&  -\frac{1}{2}\rho C_{D,t}\pi  D \int_{0}^{L} (\dot{y}_{C}^k+\dot{\theta}_2^kr\cos\theta_2^k )  (\dot{x}_{C}^k\cos\theta_2^k - \\
     &\dot{y}_C^k \sin\theta_2^k )^2  (|\vec{V}_2^k(r)|)^{-1} dr,
  \end{split} 
  \end{equation}
  where $(|\vec{V}_2(r)|)^{-1}=[\dot{x}_{C}^2+\dot{y}_{C}^2+\dot{\theta_2}^2 r^2 +2\dot{\theta_2}r (\dot{x}_{C}\sin\theta_2- \dot{y}_{C}\cos\theta_2)]^{-\frac{1}{2}}$, such that $\vec{V}_2$ is the velocity of cable-segment 2.
   The above drag force components are obtained using Eqns. \eqref{eqdrag2} and \eqref{eqdrag3} and the integrations are performed using the numerical integration method. The final expression for the moment balance equation at point A is
   \begin{equation}\label{eqmdrag2} \begin{split}
    & \textstyle\sum M_{d,A}^k- (m_2\ddot{x}_{G2}^k -D_{n,x,2}^k - D_{t,x,2}^k + F_{x,3}^k)l_1 \sin\theta_{1}^k  - \\ &  (m_2\ddot{y}_{G2}^k
      - D_{n,y,2}^k- D_{t,y,2}^k +  F_{x,3}^k)l_1 \cos \theta_{1}^k  =  I_1\ddot{\theta}_{1}^k+\frac{m_1l_1}{2} \\ & (\ddot{y}_{A}^k \cos\theta_{1}^k  - \ddot{x}_{A}^k \sin\theta_{1}^k ).
 \end{split}\end{equation}
$\textstyle\sum M_{d,A}^k=f(\theta_{1}^k,\dot{\theta}_{1}^k,\theta_{2}^k,\dot{x}_{A}^k,\dot{y}_{A}^k,C_{D,n},D,l_1,m_1)$ is obtained using Eqn. \eqref{eqmdrag}. The second-order ODE of $\ddot{\theta}_{1,k}$ from Eqn. \eqref{eqmdrag2} is solved using the numerical method to obtain the yaw orientation of cable-segment 1 at instant $k$.

 The moment balance equation at node C for the cable-segment 2 is
 \begin{equation} \begin{split} \label{odecable2}
     \textstyle\sum M_{d,C}^k-\textstyle\sum M_{rtn,C}^k =&   I_2\ddot{\theta}_{2}^k+\frac{m_2l_2}{2}(\ddot{y}_{C}^k \cos\theta_{2}^k - \ddot{x}_{C}^k \\ & \sin\theta_{2}^k ),
 \end{split}\end{equation}
 where $\textstyle\sum M_{d,C}^k=f(\theta_{2}^k,\dot{\theta}_{2}^k,\theta_3^k,\dot{x}_{C}^k,\dot{y}_{C}^k,C_{D,n},D,l_2,m_2)$ is the moment generated at node $C$ due to drag force acting on segment CD which is obtained using Eqn. \eqref{eqmdrag}. $\textstyle\sum M_{rtn,C}^k$ is the moment generated at node $C$ due to reaction force from sensor-array EF, given as 
     \begin{equation} \label{eqmdrag3} \begin{split}
     \textstyle\sum M_{rtn,C}^k = & (m_2\ddot{x}_{G3}^k -D_{n,x,3}^k - D_{t,x,3}^k)l_2 \sin\theta_{2}^k -(m_2\ddot{y}_{G3}^k \\ & -D_{n,y,3}^k - D_{t,y,3}^k)l_2 \cos \theta_{1}^k,
     \end{split}\end{equation}
where $D_{n,3}^k = (D_{n,x,3}^k,D_{n,y,3}^k)$ and $D_{t,3}^k = (D_{t,x,3}^k,D_{t,y,3}^k)$  are the $(x,y)$ components of normal and tangential drag force, respectively, acting on the sensor-array EF. 
They are the function of $(D_{n,3}^k, D_{t,3}^k)=f(\theta_{3}^k,\dot{\theta}_{3}^k,\dot{x}_{E}^k,\dot{y}_{E}^k,C_{D,n,array},C_{D,t,array},D_2,l_3,m_3)$ and can be obtained using Eqn. \eqref{eqdragcomp}. The second-order ODE of $\ddot{\theta}_{2,k}$ given in Eqn. \eqref{odecable2} is solved to obtain the yaw orientation of cable-segment 2 at any instant $k$.

The sensor-array end is free, so there is no reaction force at the end. The moment balance equation at node E of sensor-array is given as
 \begin{equation} \label{odearray}
     \textstyle\sum M_{d,E}^k =  \dfrac{m_3l_3^2}{3}\ddot{\theta}_{3}^k+\frac{m_3l_3}{2}(\ddot{y}_{E}^k \cos\theta_{3}^k - \ddot{x}_{E}^k \sin\theta_{3}^k ),
 \end{equation}
 where $m_3$ and $l_3$ are the mass and length of the sensor-array. $\textstyle\sum M_{d,E}^k=f(\theta_{3}^k,\dot{\theta}_{3}^k,\dot{x}_{E}^k,\dot{y}_{E}^k,C_{D,n,array},D_3,l_3,m_3)$ is obtained using equation (15). The orientation of sensor-array $\theta_{3}^k$ is obtained by solving the second-order ODE of $\ddot{\theta}_3^k$ given in Eqn. \eqref{odearray}.
 
 The yaw orientation of the discretized cable segment and sensor-array \emph{i.e.} $\theta_1,\theta_2,\theta_3$ are obtained by solving the $2^{nd}$ order ODEs from Eqns. \eqref{eqmdrag2}, \eqref{odecable2}, and \eqref{odearray}, respectively, which are then utilized to obtain the position, velocity and acceleration of each node using kinematic relations from Eqn. \eqref{eqkinematics}. 
 The boundary condition at the cable's top end \emph{i.e.} position, velocity and acceleration of node A at $k^{th}$ time instant, are the same as those of the ship, which are known functions of time.

\section{Simulation Results}
\subsection{Scenario}
\begin{figure}
				\includegraphics[scale = 0.65]{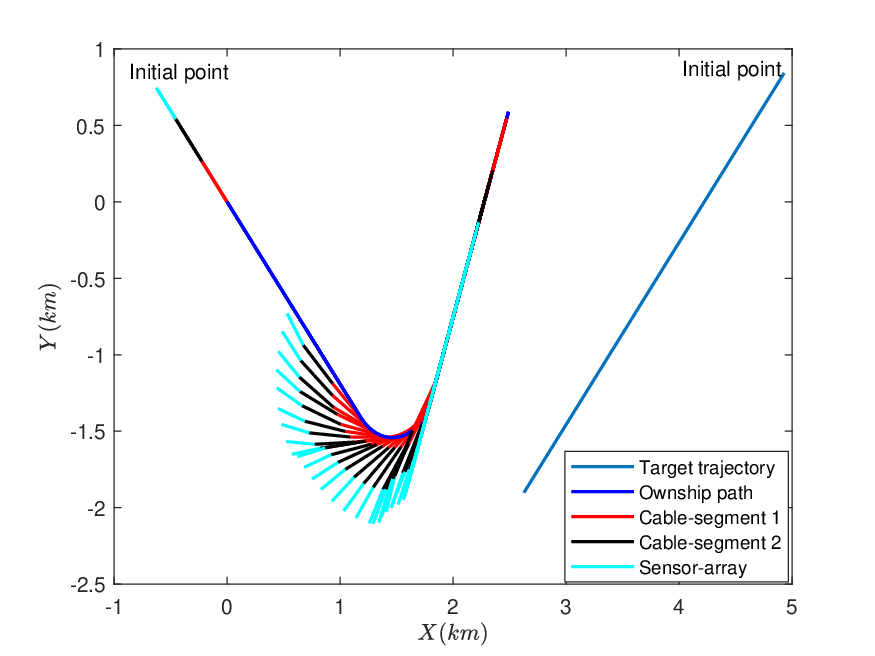}
				\caption{Dynamics of towed cable- sensor array system.}
				\label{result1}
\end{figure}

	\begin{table}
				\caption{Physical properties of towing cable sensor-array system \cite{guo2021numerical}.}
				\begin{center} 
					\begin{tabular}{ p{4cm} p{1.5cm}p{1.5cm} } 
						\hline
						\textbf{Parameters} & \textbf{Towing cable} & \textbf{Sensor-array}\\
						\hline
                    Length $(m)$ & 723 & 273.9\\
                    Diameter $(m)$ & 0.041 & 0.079\\
                    Mass/length $(Kg/m)$ & 2.33 & 5.07 \\
                    $C_{D,n}$ & 2 & 1.8\\
                    $C_{D,t}$ & 0.015 & 0.009\\
						\hline
					\end{tabular}
                 \label{parameters}
				\end{center}
			\end{table}
The surface vessel is moving with constant velocity motion with a cable sensor-array system connected to it through a spherical joint. 
The dynamics of the towed cable sensor-array system is simulated for a scenario involving the engagement of a target and an own-ship, which is popularly used in many existing TMA problems \cite{radhakrishnan2018gaussian},\cite{kumar2022tracking},\cite{singh2022passive}.
The initial coordinate of the vessel is at origin $(0,0,0)km$ moving with a constant velocity of $5 \ knots$ for a period of $30 \ min$. The vessel starts at a course of $140^o$ with respect to $y-axis$ and continues moving at that path for a period of first $12 \ min$. Then, the vessel performs a manoeuvre from $13^{th}$ to $17^{th}$ time steps with a constant turn rate of $30^o \ per \ min$ and ends up at $20^o$ course with respect to y-axis. The vessel continues to move at a course of $20^o$ with respect to y-axis from $18^{th}$ to $30^{th}$ minute. 
The course followed by the target and ownship is shown in Fig. \ref{result1}.
%The initial position of the target is at $(4.9286, 0.8420,0)km$. It is moving with a constant velocity of $4 \ knots$ at a straight course of $-140^o$ \emph{w.r.t} $y$-axis, for a period of $30 \ min$. 
The physical properties of the towing cable-array system are given in Table \ref{parameters}, which is taken from Ref. \cite{guo2021numerical}. The density of fluid is taken to be $1000\ kg/m^3$.
\begin{figure}
				\includegraphics[scale = 0.65]{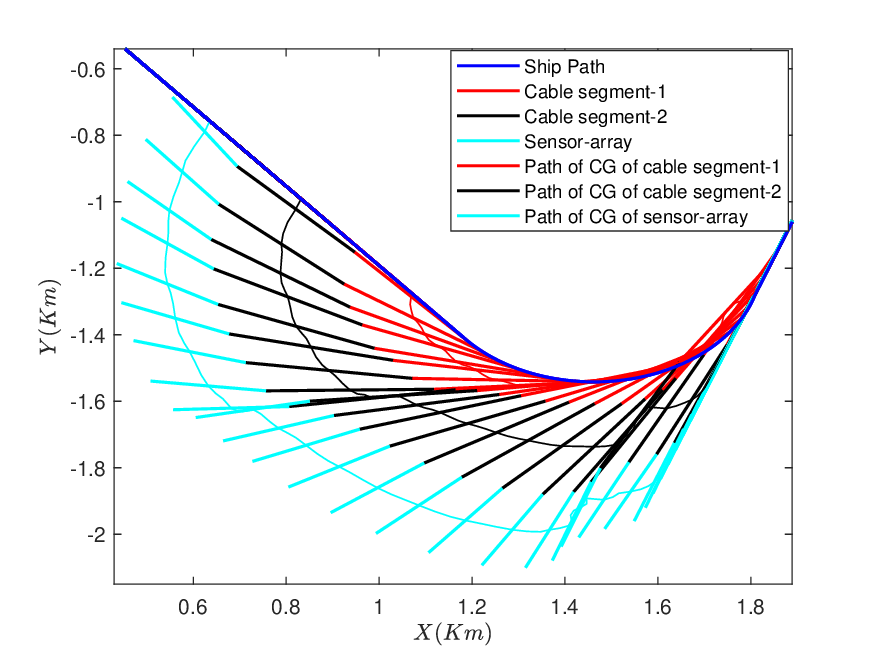}
				\caption{Enlarged view showing the dynamics of towed cable-sensor array system during a manoeuvre of the ship.}
				\label{zoomcablearray}
\end{figure}
\subsection{Results and Discussion}
The depth attained by towed cable sensor-array system is determined using Eqn. \eqref{eqdepth}. The steady-state orientation \emph{w.r.t.} sea-surface \emph{i.e.} $xy$ plane is $5.64^o$, which is relatively small, such that the TCSAS can be considered to be floating on the sea surface.
The simulation is performed using MATLAB R2018a software. The second order ODEs given by Eqns. \eqref{eqmdrag2}, \eqref{odecable2}, and \eqref{odearray} are solved simultaneously using the Runge-Kutta $4^{th}$ order method \cite{press2007numerical} to obtain the orientation \emph{i.e.} $\theta_1$, $\theta_2$, and $\theta_3$. Subsequently, these orientations are inputted into the kinematic Eqn. \eqref{eqkinematics} to obtain the position, velocity and acceleration of the TCSAS at each instant $k$. 
The solution for the second order ODE requires the initial condition. 
The initial yaw for the two cable segments and one sensor-array are taken to be same as that ship initial yaw orientation, and the initial yaw rate is taken to be zero, such that the TCSAS is aligned with the surface ship initially.
The definite integral involved in the calculations of moment due to forces, reaction force, and drag force from Eqns. \eqref{eqmdrag}, \eqref{eqreaction}, and \eqref{eqdragcomp}, respectively are solved using the numerical integration method called Gauss-Legendre five point rule \cite{press2007numerical}. For example, the integration of moment due to drag force is approximated using the points $\zeta_i$ and their corresponding weights $w_i$, given as $\int_{0}^{L} M_{d,X} dr = \textstyle\sum_{i=1}^{n} w_i M_{d,X}(\zeta_i)$.
The dynamics of 2 cable-segment and sensor-array for the simulated ship's trajectory is shown in Fig. \ref{result1}. The red, black and cyan color plots show the path traced by cable segments and sensor-array, respectively at each time instant.
Initially, it is evident that the path of the towed body aligns with the ship's trajectory during straight-line motion. However, upon initiation of the ship's manoeuvre, the towed body deviates from its original path. Following the completion of the manoeuvre at the $17^{th}$ time step, the towed body gradually stabilizes due to its inertia and reaches a steady-state position after some delay.
The enhanced view of the dynamics of the towed array during the manoeuvring of ship is shown in Fig. \ref{zoomcablearray}, which shows the path traced by the centre of gravity (CG) of the cable segment and sensor array. 
The location of CG of sensor array is utilized to perform the TMA efficiently, especially during the manoeuvring of the own-ship.
\section{Conclusions}
This paper develops a dynamic model of a towed cable sensor-array system using the lumped mass approach.  The developed model when solved at each node, provides the location and orientation of the towed array sensor \emph{w.r.t.} own-ship motion. The coordinates of CG of the sensor-array are to be fed to state estimation model to perform an accurate TMA of an underwater target, especially during manoeuvres. As a future work, it is intended to develop a generalized dynamics model which can consider the towed system as a combination of any user defined number of rigid bodies. The position of the towed array obtained out of the developed model shall be fed to a state estimation algorithm, which will recommend using optimal control or potential field algorithms to steer a tracking vessel into a dynamically determined preferred tracking position. With the use of heuristics on ship movement and detection parameters, a means to safely track very quiet submerged targets could be derived.

%The second-order ODE obtained is solved using Runge-Kutta fourth order numerical method
%\appendix
 %The Gauss-Legendre five-point rule \cite{press2007numerical} is employed to perform the definite integral.
 %   Here, 5 sample points $\zeta_i$ are used, and their associated weights are $w_i$. $\zeta_i$ are the roots of $n^{th}$ order Legendre polynomial. The Legendre polynomial is $(n+1)P_{n+1}(\zeta)=(2n+1)\zeta P_{n}(\zeta)-nP_{n-1}(\zeta)$. $w_i$ are the quadrature weights given as $w_i = \dfrac{2}{(1-\zeta_i^2)[\dot{P}_n(\zeta_i)]^2}$. The range of integration is scaled from $(0, \ L)$ to $(-1,\ 1)$, enabling the integration to be approximated using the points $\zeta_i$ and weights $w_i$. Mathematically, this is expressed as: \begin{equation} \label{eqmoment2}
  %      \int_{0}^{L} M_{d,X} dr = \textstyle\sum_{i=1}^{n} w_i M_{d,X}(\zeta_i).
  %  \end{equation}

	\end{document}